\documentclass[11pt]{article}
\usepackage{moriond,epsfig}
\pdfoutput=1 

\def\Journal#1#2#3#4{{#1} {\bf #2}, #3 (#4)}

\def\PLB{{\em Phys. Lett.}  B}

\def\PRD{{\em Phys. Rev.} D}

\def\be{\begin{equation}}
\def\ee{\end{equation}}
\def\bea{\begin{eqnarray}}
\def\eea{\end{eqnarray}}

\def\ttbar{t\bar{t}}
\def\met{E_{\rm T}^{\rm miss}}
\def\mt{M_{\rm T}}
\def\pt{p_{\rm T}}
\def\re{{\rm Re}}

\begin{document}
\vspace*{4cm}
\title{Recent results on Top quark Physics with the ATLAS and CMS experiments}

\author{ P. Silva,~{\em on behalf of the ATLAS and CMS collaborations} }

\address{CERN, CH-1211 Gen\`eve 23, Switzerland and \\ LIP, Av. Elias Garcia 14 - 1$^o$ 1000-149 Lisboa, Portugal}

\maketitle\abstracts{
An overview of the most recent results on top quark physics obtained
using proton-proton collision data collected with the ATLAS and the CMS experiments at 7 TeV center-of-mass energy is given. 
Measurements for inclusive and differential top quark pair or single top quark production in different final states are reviewed.
Top properties such as W helicity in top decays, electric charge, charge asymmetry and spin correlations of top quark pairs,
among others have also been measured by the LHC experiments.
All the measurements are found to be consistent with the Standard Model predictions with a good level of accuracy.
The latest results in the measurement of the top quark mass at the LHC are also presented and discussed.
}

%
%
\section{Introduction}
\label{sec:intro}

The Large Hadron Collider (LHC) has performed extraordinarily well throughout the year of 2011
providing both the ATLAS~\cite{Aad:2008zzm} and the CMS~\cite{Chatrchyan:2008aa} experiments with over a million top quarks each.
The top quark is one of its kind - it has a mass close to an atom of Gold
and a large width which makes it decay before it transverses any significant distance 
to disturb the color field giving rise to non-perturbative effects such as its fragmentation and hadronization~\cite{bigi}. 
Therefore all properties of the ``naked'' quark are preserved in the decay chain of a top quark.
Interestingly there are also many unknowns in the top quark sector, namely on the role it plays in the electroweak symmetry breaking mechanism (EWKSB)
given that it is the heaviest fundamental particle and that its mass is close to the EWKSB scale.
The ATLAS and CMS experiments have eagerly analysed the data they have collected seeking to understand better the top quark production
and its basic properties such as: mass, charge, branching ratios and kinematics of the decay products.
In these proceedings we can hardly condense the rich top quark programme at the LHC
so we highlight the most recent results obtained by the two collaborations
and refer the interested reader to the web pages with the public results~\cite{atlaspublic,cmspublic}.

\section{Top quark production}
\label{sec:xsecs}

\subsection{Top pair production}
\label{subsec:ttbarxsec}

At the LHC the top pairs are mostly ($\approx$90\%) produced through gluon-gluon fusion.
The measurement of the $\ttbar$ cross section has been carried out in all the different final states
which result from the combinatorics of the $W$ boson decays, since the top quark decays mostly in the $t\rightarrow Wb$ channel.
The $\ttbar$ decay channels comprise therefore fully-hadronic (46\%), lepton+jets (45\%) and dileptonic (9\%) final states.
The most precise measurement is obtained in the lepton+jets channel due to both its high statistics
and the strategy followed by both experiments.

In the lepton+jets channel  the main backgrounds are due to QCD multijets, W+heavy flavour.
While the transverse energy flux in QCD multijets processes is expected to be well balanced, 
constraining these processes in the lower region of the missing transverse energy ($\met$),
and transverse mass ($\mt$ 
\footnote{The transverse mass of a single lepton event with $\pt^{\ell}$ and the $\met$ is defined as $\mt=\sqrt{2(1-\cos\delta\phi)p_T^l \met}$,
where $\delta\phi$ is the angle between the lepton transverse momentum and the $\met$ direction.}
) spectra, this is no longer the case for $W$ events.
Therefore, given the similarities of the $W$+jets background and the signal, 
the strategy adopted by both the ATLAS and CMS experiments is to analyse different categories of events
according to the jet multiplicity and the number of $b$-tagged jets, i.e. identified as $b$ jets.
This procedure has the advantage of being able to constrain the actual contamination from background processes
which are expected to have lower jet multiplicity than the signal (i.e. $<$4 jets) and lower heavy flavour content (i.e. $b$ jets).
The cross section is extracted after fitting  either a simple and robust variable as the mass of the secondary vertex for the jets (CMS~\cite{Sadia})
or a multivariate discriminator based on the kinematics of the event (ATLAS~\cite{ATLAS:11121top})  to the different event categories.
The fit takes into account not only the normalisation of the background processes but also
how it can be affected by the different systematic uncertainties such as jet energy scale,
$b$-tag or mistag efficiencies, the contamination from initial/final state radiation (ISR/FSR)
and the factorisation and renormalization scales 
use to model the signal and some of the backgrounds (i.e. $Q^2$ scale).
The relative uncertainty in the measurement of the $\ttbar$ cross section by the ATLAS collaboration is $\approx$7\% 
and it is dominated the uncertainty in modelling of the signal component and by the measurement of the luminosity.

In the remaining channels the $\ttbar$ production cross section has been carried out mostly through
counting experiments with the exception of: the fully hadronic channel, which has used a fit to the
distribution of the reconstructed top mass with a kinematics fit (CMS) 
or the $\min \chi^2$ assuming $m_{top}=$172.5~GeV/c$^2$ found from all possible jet combinations in the event (ATLAS),
and the $\tau$ channels where a multivariate analysis is used (ATLAS).
Overall the results are compatible with the theoretical predictions 
but its uncertainties tend to be larger with respect to the measurement in the lepton+jets channel
due to systematic effects such as jet energy scale or background estimations.

\begin{figure}
\centering
\includegraphics[height=2.8in]{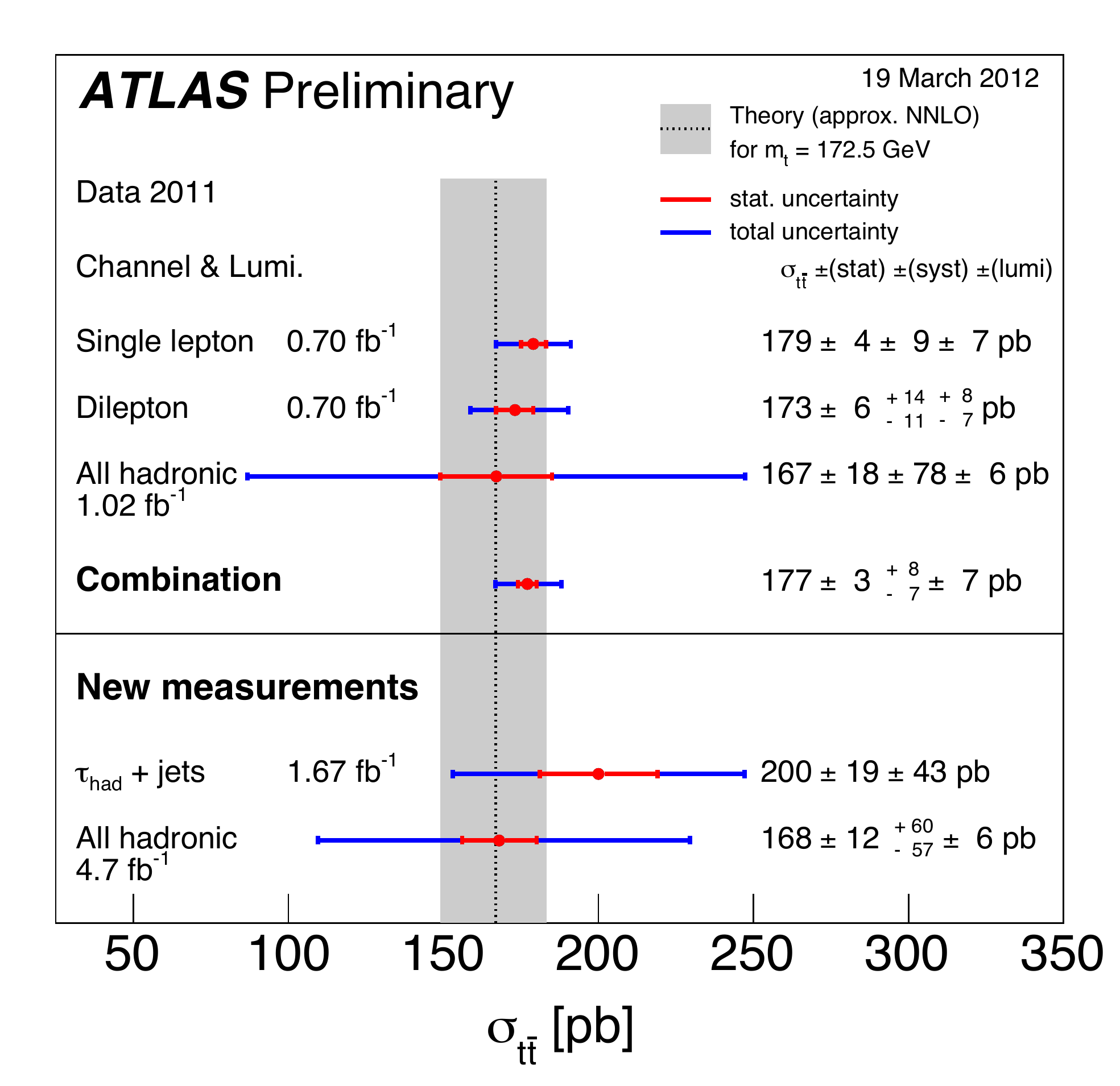}~~~
\includegraphics[height=2.8in]{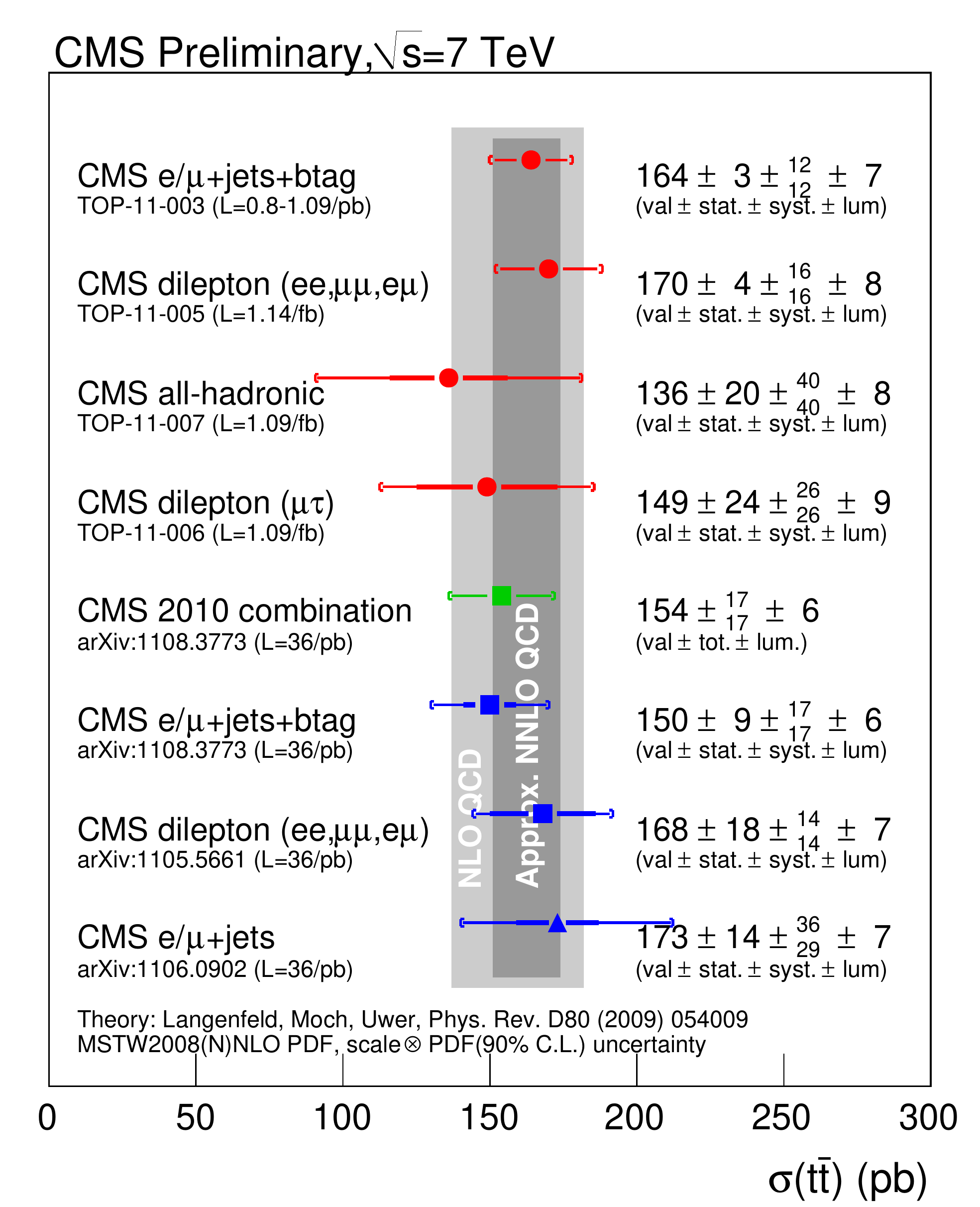}
\caption{Summary of the $\ttbar$ cross section measurements at $\sqrt{s}=$7~TeV performed by the ATLAS ({\em left}) and CMS ({\em right}) collaborations.
Different theoretical predictions are shown as represented as vertical bands.}
\label{fig:ttbarxsec}
\end{figure}

Figure~\ref{fig:ttbarxsec} summarises the results obtained in the different channels and the final combination.
The total uncertainty attained by each experiment has now surpassed the theoretical uncertainty at approx. NNLO.

Besides inclusive $\ttbar$ measurements the associated production with a photon has
also been measured by the ATLAS collaboration and the results are found to be in good agreement with the SM predictions:
$\sigma(\ttbar\gamma)=2.0 \pm 0.5_{\rm stat} \pm 0.7_{\rm syst} \pm 0.08_{\rm lumi}$~pb~\cite{ATLAS:11153top}.

With the large statistics sample acquired in 2011 the inclusive measurements
were also expanded to measure differential cross sections such as $\ttbar$+N~jets, 
$p_T^{\ttbar}$, $p_T^{top}$, $M_{\ttbar}$, among others.

\begin{figure}
\centering
\includegraphics[height=2.5in]{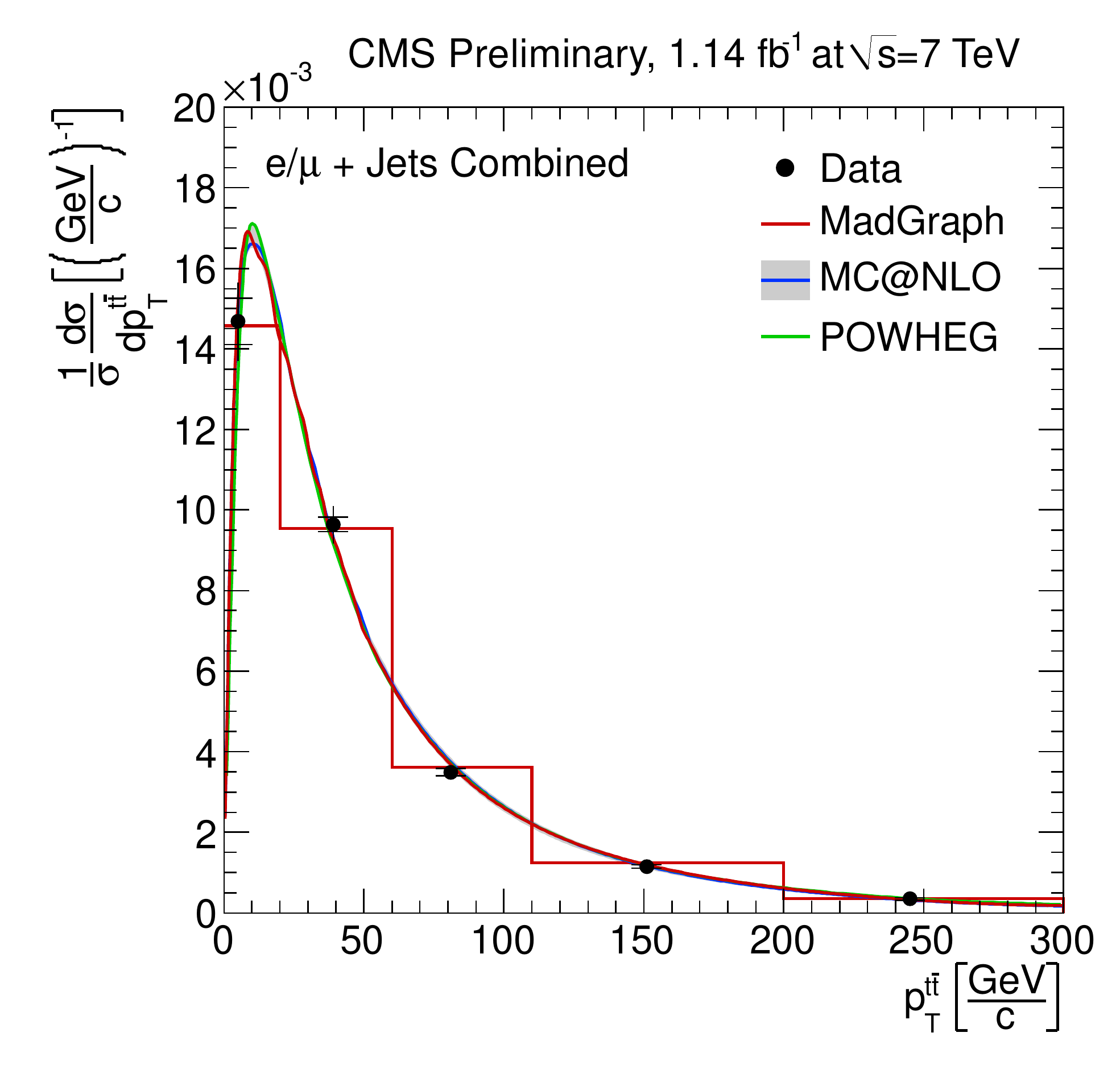}~~~
\includegraphics[height=2.3in]{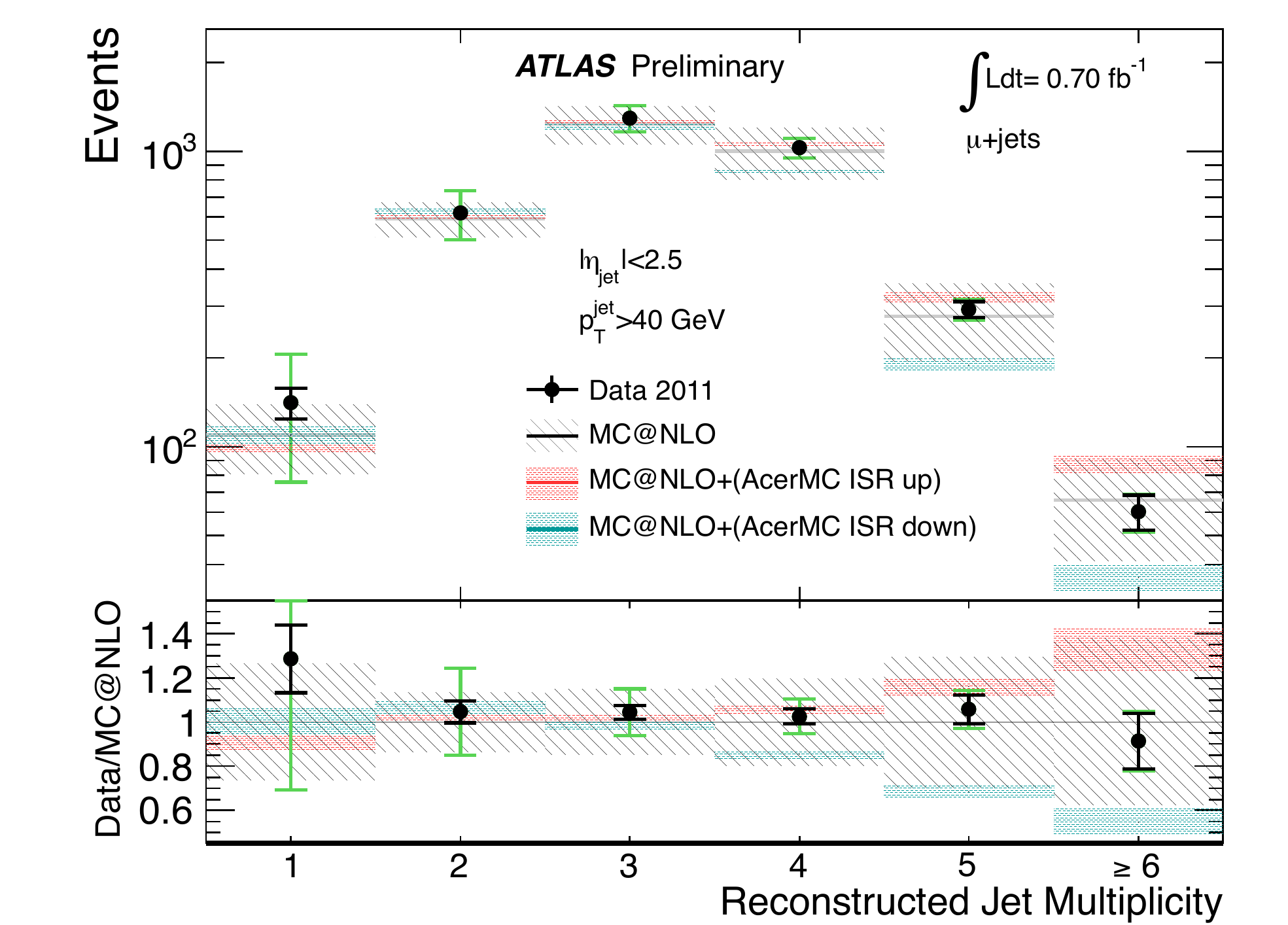}
\caption{Differential cross section measurements in the lepton+jets channel.
{\em Left}: The transverse momentum of the top quark pair obtained after unfolding is compared to different MC generators.
{\em Right}: The jet multiplicity distribution after background subtraction is compared to the nominal and varied ISR showering predictions.}
\label{fig:ttbardiffxsec}
\end{figure}

Differential measurements were carried out in the lepton+jets and dilepton channels
after the reconstruction of the $\ttbar$ kinematics~\cite{CMS:11013top}.
In the lepton+jets this is achieved after choosing the combination of jets which yields the best kinematical fit.
In the dilepton channel due to the presence of two neutrinos in the final state, and therefore unconstrained kinematics,
the solutions for the kinematics are found scanning $m_{top}$ in the [100,300]~GeV/c$^2$
and compared with the simulated expectations for the neutrino kinematics.
The solution with larger probability which makes use of the highest number of $b$-tagged jets is used.
These arbitrations in the choice of the kinematics may
lead to misassignment of the objects or to event rejection in case the algorithm fails to find a valid solution.
These effects lead to an additional smearing of the reconstructed kinematics (on top of the detector resolution effects).
This effect is minimised by unfolding the reconstructed kinematics to parton level.
A Singular Value Decomposition (SVD) technique is applied and the widths of the bins ($\Delta^i_x$) of each kinematical variable
are chosen in order to keep the purity and stability above 50\%.
The differential cross section is measured after background subtraction and unfolding the observed value.
\footnote{The differential cross section with respect to a kinematical variable $x$
is usually expressed as $1/\sigma \cdot d\sigma^i / dx  = 1/\sigma \cdot (N_{data}-N_{background})/ \Delta^i_x\varepsilon^i\mathcal{L}$
where $\varepsilon^i$ is the efficiency correction for the i$^{th}$ bin and $\mathcal{L}$ is the total integrated luminosity.}
.
Overall the agreement between the unfolded data and the simulation is remarkable 
but the uncertainty attained is not yet at the level where one signal model can be preferred among the ones studied.
One distribution of particular interest is $\pt^{\ttbar}$ and it is shown in Fig.~\ref{fig:ttbardiffxsec} ({\em left}).

The measurement of $\ttbar$+N~jets is correlated with the $\pt^{\ttbar}$ spectrum
and it assesses from data how well the recoil of the $\ttbar$ system 
and the modelling of ISR are predicted theoretically and by the simulation.
The lepton+jets channel has been used to study the associated production with $\ttbar$ with extra jets~\cite{ATLAS:11142top}.
After background subtraction the jet multiplicity spectrum is compared to the simulated prediction using Acer MC~\cite{ACER}.
The nominal prediction is obtained after performing the parton showering with the Pythia generator~\cite{PYTHIA}
and can be varied in two ways by enhancing or suppressing: i) the ISR branching above the coherence scale;
ii) the $\alpha_{\rm QCD}$ evolution scale of the ISR (effect is proportional to $\Lambda^{-1}_{ISR}$).
The measurement is dominated by the uncertainties in the jet energy scale and in the subtraction of the background.
Even if compatible with the nominal prediction for the ISR in $\ttbar$ events
it has not yet attained the necessary precision to constrain further the modelling of the production of extra jets in this sample.

\subsection{Single top production}
\label{subsec:singletopproduction}

Single top quarks can be produced through the s- and t-channels and in association with a $W$ boson.
Table~\ref{tab:stopxsec} summarises the current status of the
measurement of the single top production cross section at the LHC 
~\cite{CMS:11021top,CMS:11022top,Collaboration:2012ux,ATLAS:11104top,ATLAS:11118top}
and compares the results with the theoretical predictions
~\cite{PhysRevD.83.091503,PhysRevD.81.054028,Kidonakis:2010ux}
. 
With the exception of the s-channel which will require larger amount of integrated luminosity to be observed,
the agreement between experiment and theory for the single top production cross section is remarkable.
In the following we summarise briefly the strategies followed to extract these cross sections from data.

\begin{table}[t]
\caption{Summary of the experimental and theoretical cross sections for single top production in the different channels.
See text for references.}
\label{tab:stopxsec}
\vspace{0.4cm}
\begin{center}
\hspace*{-0.6cm}
\begin{tabular}{llll}
\hline\hline
\vspace{0.2cm} {\bf Channel} & $\sigma_{\rm t}~(\rm pb)$                               & $\sigma_{\rm tW}$ (pb)                                     & $\sigma_{\rm s}$ (pb) \\\hline
\vspace{0.2cm} ATLAS   & $83\pm4({\rm stat})^{+20}_{-19}({\rm syst})$                  & $14.4^{+5.3}_{-5.1}({\rm stat})^{+9.7}_{-9.4}({\rm syst})$ & $<26.5$ @ 95\% CL \\
\vspace{0.2cm} CMS     & $70.2\pm5.2({\rm stat})\pm10.4({\rm syst})\pm3.4({\rm lumi})$ & $22^{+9}_{-7}({\rm stat}\oplus{\rm syst})$                 & - \\
\vspace{0.2cm} Theory  & $64.57^{+2.09}_{-0.71}~^{+1.51}_{-1.74}$                      & $15.74\pm0.40^{+1.10}_{-1.14}$                             & $4.63\pm0.07^{+0.19}_{-0.17}$  \\
\hline\hline
\end{tabular}
\end{center}
\end{table}

The dominant production mode is the $t$-channel and it is characterised by
one central isolated lepton accompanied by $\met$, a $b$-jet and a forward high $\pt$ recoiling jet.
Due to the proton PDFs we expect that the ratio between single top and single anti-top produced through the $t$-channel to be $\approx$1.9.
The main backgrounds to this channel are due to QCD multijets and W+heavy flavour production. 
The first type of background can be controlled in a sideband where the lepton is non-isolated and
a template for the $\met$ or $\mt$ distribution can be derived and used as input to fit in the signal region.
The second type of background is almost irreducible and can at most be constrained
(by splitting the data sample in different categories according to the jet and $b$-tag multiplicities, or failing
a requirement on the reconstructed top quark mass from the lepton-$\met$-$b$-jet system)
or partially discriminated (with a multivariate analysis).
Both approaches have been followed:
ATLAS measures the cross section by fitting the distribution of a multivariate discriminant constructed with a neural network
and the most recent result, from CMS,
uses a fit to a simple and robust variable - the pseudo-rapidity of the recoil jet -
to extract the production of single top in the $t$-channel.
The result is used to measure the CKM matrix element $V_{tb}$
and the result obtained is: $|V_{tb}|=$1.04 $\pm$ 0.09 (exp) $\pm$ 0.02 (th) (CMS).

Both experiments are close to find evidence for single top production in the $tW$-channel using dilepton events.
The definition of this channel overlaps partially with the $\ttbar$ process when one of the top quarks is virtual.
The distinctive feature of the $tW$ channel is the presence of a single $b$-jet in the majority of the events
and the fact that at LO the the decay products of the top and the $W$ boson balance each other.
Both experiments extract the $tW$ cross section from a likelihood fit for counts of events with
$n$-jets and $k$ $b$-tags. Even if the signal is absent from high multiplicity events these control regions
help further in constraining the $\ttbar$ contribution.

The rarest, and nevertheless most interesting of the channels, is the $s$-channel
which is expected to be sensitive to new physics. Although not yet observed, the ATLAS experiment has 
put upper limits on the cross section value from a kinematics analysis.
The S/B$^{1/2}$ expected is 0.98.

\section{Top quark properties}
\label{sec:topproperties}

With large statistics samples many properties can be measured accurately.
The main challenge is to constrain the systematic uncertainties which in many case 
dominate the final uncertainty. Part of these uncertainties can indeed be constrained from data
as we have highlighted in the previous section: ISR/FSR and Q$^2$-scale are some examples.

\begin{figure}
\centering
\includegraphics[height=2.7in]{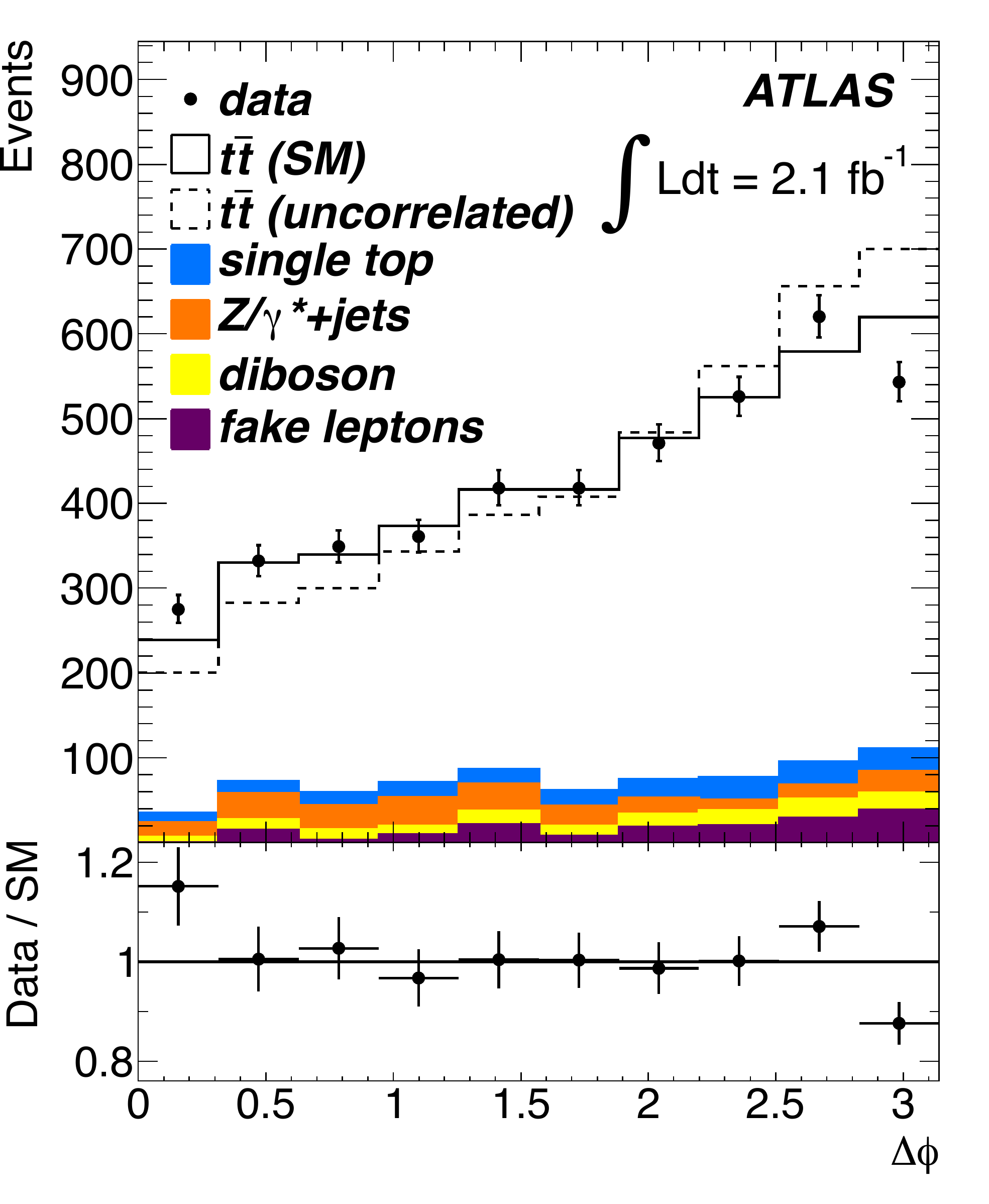}~~~
\includegraphics[height=2.9in]{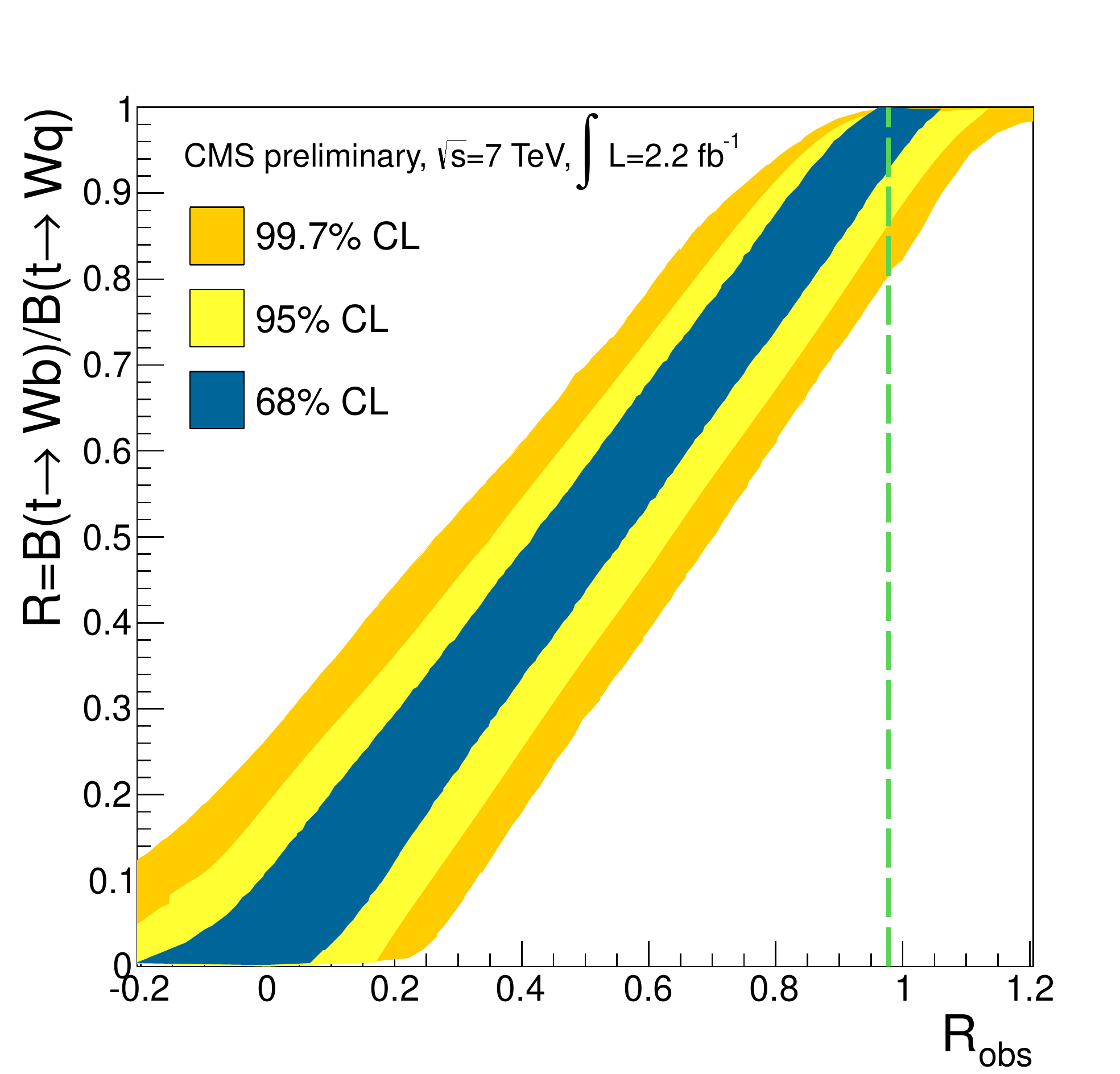}
\caption{
{\em Left}: Difference in the azimuthal angle of dilepton $\ttbar$ candidates. The data observed is compared with the SM prediction (signal+ background)
and with a possible scenario where the $\ttbar$ pair is produced without any spin correlation.
{\em Right}: Limit bands at different CL on ${\cal R}$ with the measured value overlaid as a dashed line.
The favoured values of ${\cal R}_{\rm obs}$ obtained for different CL (horizontal axis) are shown for each true value of ${\cal R}$ (vertical axis).}
\label{fig:topprops1}
\end{figure}

Information on spin correlation of the $\ttbar$ system is preserved by the decay products.
It is found that the degree of information is maximal in the lepton and down-type quark kinematics.
The measurement can be performed in two alternative ways:
i) reconstructing the full kinematics and studying the $\ttbar$ system in a specific reference frame
(usually the helicity basis where the system is at rest or a maximal frame which is defined event by event
from the top kinematics with respect to the beam-line);
ii) using a simple robust variable reconstructed from a high purity final state and translate it to a reference frame.
The second strategy was pursued by the ATLAS experiment to find evidence for spin correlations in $\ttbar$ production~\cite{ATLAS:2012ao}.
The difference in the azimuthal angle of the dilepton candidates is fit using templates for the predictions
within the SM and in the absence of correlations. The distribution is shown in Fig.~\ref{fig:topprops1} ({\em left})
and it depicts both the high purity of the sample and the evidence for correlation of the two leptons.
The result of the fits is in good agreement with the SM prediction and the asymmetries measured are:
$A_{\rm helicity}=0.34\pm0.07_{\rm stat}~^{+0.13}_{-0.09}~_{\rm syst}$ 
($A_{\rm helicity}^{\rm SM}=0.32$)
and
$A_{\rm maximal}=0.47\pm0.09_{\rm stat}~^{+0.18}_{-0.12}~_{\rm syst}$ 
($A_{\rm maximal}^{\rm SM}=0.44$).
The quoted SM predictions were computed with MC@NLO and CTEQ6.6 PDF.
The systematic uncertainty is at this point dominated by the effects of ISR and signal modelling.

The top quark charge has also been measured using the charge of the
decay products~\cite{ATLAS:11141top,CMS:11031top}.
The charge can be measured directly from the lepton from the $W$ decay and from the $b$-jet.
In the latter case two possibilities can be used: using a charge estimator based on the tracks associated to the jet
or using semi-leptonic $B$ decays which minimize charge assignments but are prone to sub-decays or oscillations
of the $B$ hadrons. Both cases can be optimised and have its efficiency measured from the abundant QCD $b\bar{b}$ production.
The events are counted in two charge possibilities (2e/3 or 4e/3) and from the observed asymmetry
an upper limit on exotic scenarios is set.
Both experiments set a $>5\sigma$ limit on exotic top quark charge production.

Both ATLAS and CMS have searched for deviations in the SM predictions for the EWK couplings of the top quarks.
Although the top quark is primarily expected to couple to a $Wb$ vertex, Flavour Changing Neutral Currents (FCNC)
may contribute because of the presence of new physics.
Anomalous $tWb$ couplings can also be enhanced by the presence of new physics.

The search for FCNC in $\ttbar$ events 
has been carried out in trilepton events where one $t\rightarrow Zq\rightarrow \ell\ell q$ is produced.
In this topology the full kinematics of the event is specified (the $\met$ is the direct signature of the escaping neutrino)
and the objects can be assigned to each top decay by using a kinematics fit 
or by requiring one of the jets to be $b$-tagged.
This search is mostly expected to be dominated by the statistical uncertainty
due to small presence of backgrounds (other than di-boson production).
The full 2011 dataset is used by the CMS experiment to set an upper
limit of ${\cal B}(t\rightarrow qZ)<0.34$~\cite{CMS:11028top}
and limits are also provided by ATLAS~\cite{ATLAS:11061top}.
Evidence for FCNC can also be sought in single top production. The signature is challenging experimentally
as $qg\rightarrow t\rightarrow W(\rightarrow \ell\nu) b$ is characterised by 
a single lepton + $b$ jet + $\met$ final state which is contaminated mostly by W+heavy~flavour production.
The ATLAS collaboration has explored the topology of the signal via a multivariate analysis.
The signal is mostly characterised by the fact that:
i) top quarks are produced at rest; 
ii) $W$ bosons are boosted (with $m_{\rm T}+\met>$60~GeV)
iii) $N(t)/N(\bar{t})\sim$4;
An upper limit is set on the production cross section $\sigma^{\rm obs}_{qg\rightarrow t}<$3.9~pb at 95\% CL 
(with an expectation of $\sigma^{\rm exp}_{qg\rightarrow t}<$2.4~pb).
Limits on FCNC decays of the top quark are set in the ${\cal B}(t\rightarrow ug)$
versus ${\cal B}(t\rightarrow cg)$ plane yielding
${\cal B}(t\rightarrow ug)<$5.7$\cdot$10$^{-5}$, if ${\cal B}(t\rightarrow cg)$=0,
and
${\cal B}(t\rightarrow cg)<$2.7$\cdot$10$^{-4}$, if ${\cal B}(t\rightarrow ug)$=0.
More details can be found in~\cite{Nedden}.

A generic estimator for deviations of the predicted ${\cal B}(t\rightarrow Wb)$ is the measurement of 
${\cal R}={\cal B}(t\rightarrow Wb)/{\cal B}(t\rightarrow Wq)$. This measurement was made from the
modelling of the $b$-tag multiplicity distribution observed in $\ttbar$ dilepton events
which are expected to be highly pure in signal.
The residual background and the model for the counted $b$-tags is derived from data taking into account
not only the $b$-tag and mistag probabilities (as measured from $QCD$ dijet events, an orthogonal sample)
but also the probability to fully reconstruct and accept in the event the decay products of a top quark decay.
The result is well in agreement with the SM prediction: ${\cal R}=$0.98$\pm$0.04 (stat+syst)
and a lower endpoint for the confidence interval at 95\% CL is set
using the Feldman-Cousins approach: ${\cal R}>$0.85
~\cite{CMS:11029top}.

The decay products of the top quark can be further analysed to look for deviations from the SM predictions. 
The polarisation of the $W$ boson is of great interest as new physics
might lead to anomalous $tWb$ couplings. Given the fact that spin
information is preserved and that the $b$ quark has an almost
negligible mass compared to the top and the $W$, the SM predicts,
through the V-A couplings, that the $W$ bosons from top quark decays
are mostly longitudinally polarised  ($F_0=$0.687$\pm$0.005) or
left-handed ($F_{\rm L}=$0.311$\pm$0.005)~\cite{Czarnecki:2010wh}.
The presence of anomalous couplings would
thus lead to deviations of the fractions of polarised $W$ bosons.
These fractions can be derived from the distribution of the angle
between the lepton from the $W$ decay and the $b$-jet from the same
top decay, evaluated in the $W$ boson rest frame. In first
approximation $\cos\theta^* \approx 4 p_b\cdot p_\ell/(m_T^2-m_W^2)-1$.
The experimentally observed distribution has to be corrected for
experimental effects such as acceptance (introduced by the selection) and resolution (mostly from
the jets). Moreover the theoretical prediction needs to take properly
into account effects such as the $Q^2$, ISR/FSR, mass of the $b$
quark. In order to interpret the observed spectrum based on general
polarisation scenarios a re-weighting procedure may be applied taking
into account all these effects (CMS~\cite{CMS:11020top})
or using dedicated samples generated with Protos~\cite{protos} (ATLAS~\cite{Collaboration:2012ky}).
ATLAS introduces further the
measurement of the asymmetries of the spectrum which are robust
estimators less prone to the effect of some of the systematic
uncertainties.
The angular asymmetries are defined in such a way that
the $F_{\rm L}$ and the $F_{\rm R}$ contributions are allowed to
cancel out:
\mbox{$A_{\pm}=[N(\cos\theta^*>z)-N(\cos\theta^*<z)]
~/~[N(\cos\theta^*)>z)+N(\cos\theta^*)<z) ]$},
where $z = \pm(1-2^{2/3})$.
Table~\ref{tab:polfracs} summarises the results
obtained for the contributions from differently polarised $W$ bosons.
A good agreement is found with respect to the SM predictions and the
results are used to set limits on anomalous $tWb$ couplings.

\begin{table}[t]
\caption{Measured fractions of longitudinally, left- and right-handed
  polarisation by the ATLAS and CMS collaborations.
 The statistical and total systematic uncertainties are shown separately.
 The limits on the  anomalous couplings (and effective operator
 coefficients) are also included when available.}
\label{tab:polfracs}
\begin{center}
\begin{tabular}{lcc}
\hline\hline
Measurement   & ATLAS                                                 & CMS \\\hline
$F_0$               & 0.67 $\pm$ 0.03 $\pm$ 0.06            & 0.57 $\pm$ 0.07 $\pm$ 0.05       \\
$F_L$               & 0.32 $\pm$ 0.02 $\pm$ 0.03            &  0.39 $\pm$ 0.05 $\pm$ 0.03        \\
$F_R$               & 0.01 $\pm$ 0.01 $\pm$ 0.04            &  0.04 $\pm$ 0.04 $\pm$ 0.04          \\\hline
$\re~V_R$ {\small ($\re~C_{33}^{\phi\phi}/\Lambda^2$)} & [-0.20,0.23] {\small ([-6.7,7.8])}  & -      \\
$\re~g_L$ {\small ($\re~C_{33}^{dW}/\Lambda^2$)}        & [-0.14,0.11] {\small ([-1.6,1.2])}  & -     \\
$\re~g_R$ {\small ($\re~C_{33}^{uW}/\Lambda^2$)}        & [-0.08,0.04] {\small ([-1.0,0.5])}  & [-0.17,0.02] ({\small [-1.9,0.2]})      \\
\hline\hline
\end{tabular}
\end{center}
\end{table}

\begin{figure}
\centering
\includegraphics[height=2.6in]{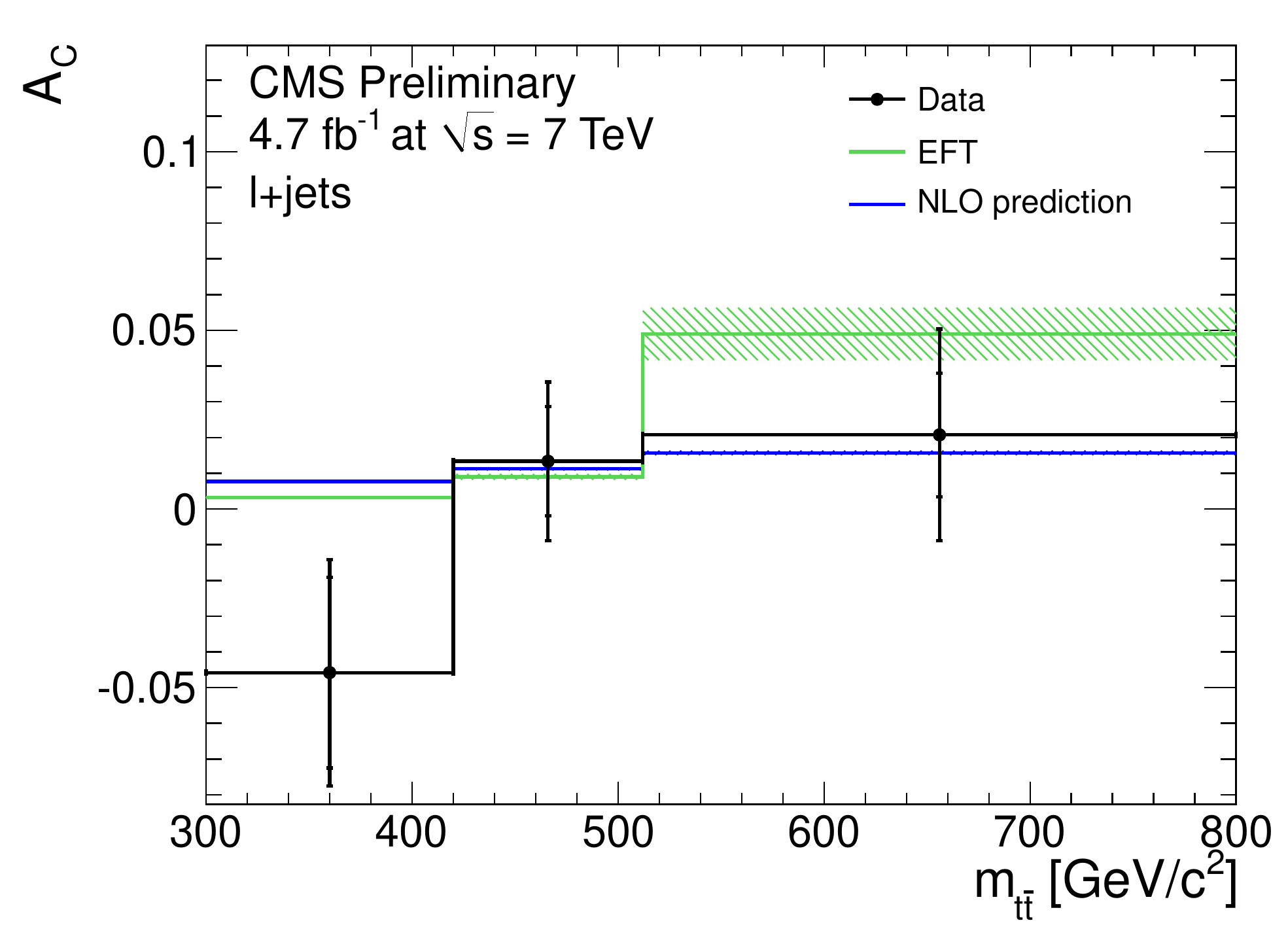}~~~
\includegraphics[height=2.5in]{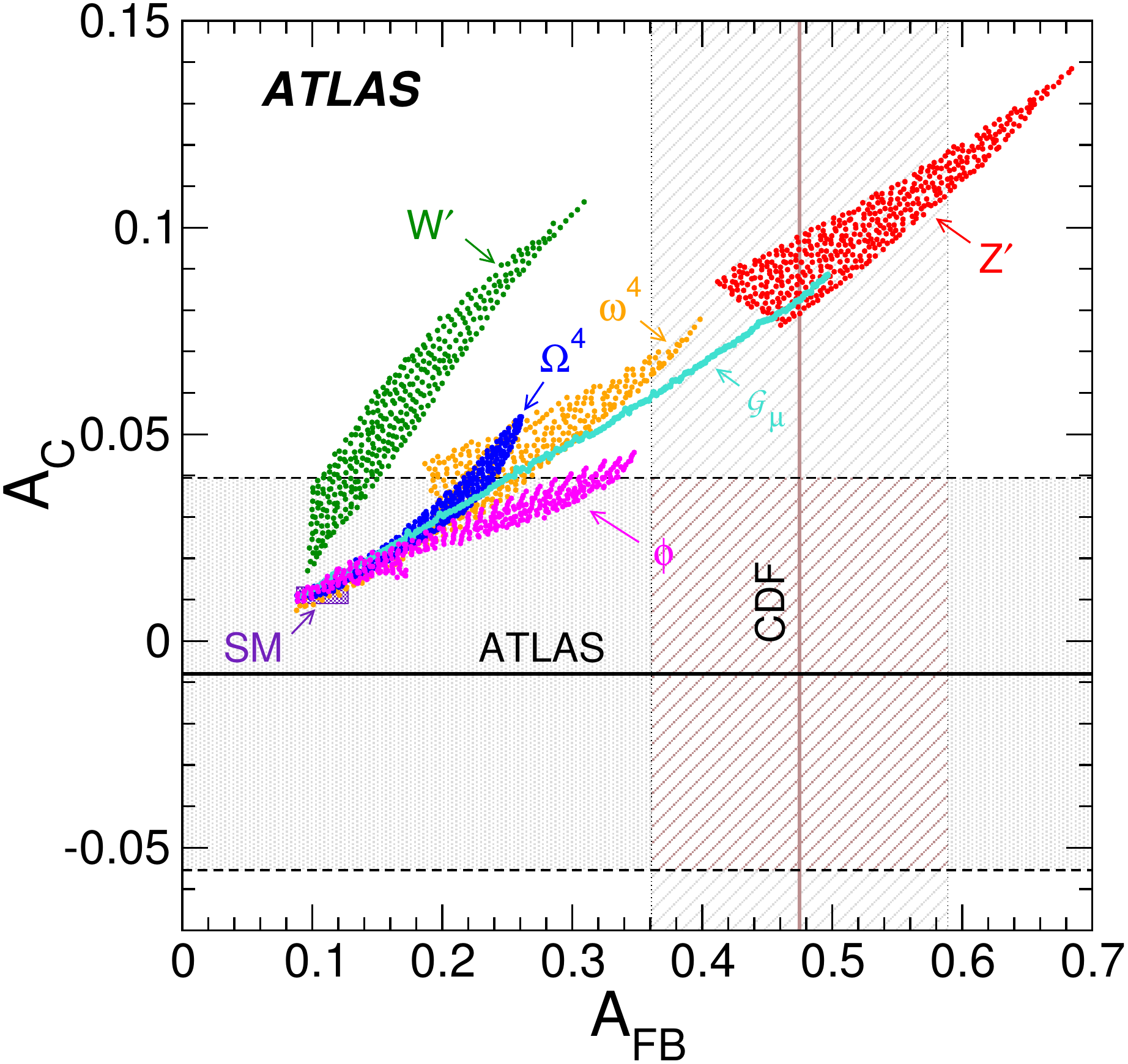}
\caption{
{\em Left}: Observed asymmetry as function of the invariant mass of the $\ttbar$ pair compared to SM predictions and to an effective field theory (see text).
{\em Right}: Limits set on BSM models from the observed charge asymmetry compared with the value measured by the CDF experiment.}
\label{fig:topprops2}
\end{figure}

We conclude the summary of the measurements of the top quark
properties with a discussion on the top pair charge asymmetry which is
expected to be highly sensitive to new physics effects~\cite{PhysRevD.84.054017}.
In $p-p$ collisions the asymmetry manifests through 
a preferential production of top quarks in the forward direction 
due to the fact that the
anti-quarks from the proton's sea tend to carry a lower momentum
fraction. 
The quantity of interest is therefore the charge asymmetry which
can be written as 
\mbox{$A_{\rm C}=(N^+-N^-)/(N^++N^-)$} 
where
$N^+$ ($N^-$) is the number of events with 
\mbox{$\Delta \eta =|\eta_{t}|-|\eta_{\bar{t}}|>0$} ($<0$)
\footnote{The events can also be counted using $\Delta
  y^2=(y_t-y_{\bar t})(y_t+y_{\bar t})=(y_t^2-y_{\bar t}^2)$ as
  alternative variable.}.
The asymmetry is measured in lepton+jets events.
In order to reconstruct the kinematics of the top and anti-top
complementary approaches are used:
i) a kinematics fit (based on a likelihood approach)
that assesses the compatibility of the observed event with the decays
of a top-antitop pair~\cite{ATLAS:2012an};
ii) the full reconstruction of the leptonically decay W and the top
which originated it (after complementing with a selected jet) followed
by the reconstruction of the second top from three selected
jets~\cite{CMS:11030top}.
In the latter case a probability is defined for each jet permutation
and the combination with larger probability is used to reconstruct
$|\Delta \eta|$ (expected to be correct in 72\% of the cases).
After background subtraction of the reconstructed events an unfolding
procedure is applied to recover the parton level kinematics.
CMS applies a regularised unfolding procedure to the data
through a generalised matrix-inversion method using 
twice as many bins for the uncorrected as for the corrected spectrum.
ATLAS uses a bayesian approach to invert the response matrix
in order to find the corrected spectrum.
Both experiments measure an asymmetry which is compatible with the NLO
prediction: $A_{\rm C}^{\rm NLO}=$0.0115$\pm$0.0006~\cite{Rodrigo}.
ATLAS measures 
$A_{\rm C}=$-0.018 $\pm$0.028$_{\rm stat}$ $\pm$0.023$_{\rm syst}$ 
using 1.04~fb$^{-1}$
while CMS measures
$A_{\rm C}=$0.004 $\pm$0.010$_{\rm stat}$ $\pm$0.012$_{\rm syst}$
using 4.7~fb$^{-1}$.
Differential measurements of the asymmetries have been carried out
namely as function of $\pt$, $y$ and invariant mass ($M_{\ttbar}$) of the $\ttbar$
pair and found to be consistent with the NLO predictions.
The dependency of $A_{\rm C}$ as function of any of these variables
has been compared to BSM models (ATLAS) or to Effective Field Theory
(CMS) and no
hints for contributions from physics beyond the standard model have
been found.
Figure~\ref{fig:topprops2} ({\em left}) shows the differential
measurement of $A_{\rm C}$ as function of $M_{\ttbar}$ as obtained by
the CMS experiment.
The ATLAS collaboration has used its results to set stringent limits
on some of the models after obtaining $A_{\rm C}$ for  $M_{\ttbar}>$450~GeV/c$^2$
as shown in Fig.~\ref{fig:topprops2} ({\em right}).

\section{Top quark mass}
\label{sec:topmass}

The mass of the top quark is a property which deserves particular attention.
It is a fundamental parameter of the EWK theory which is able to constrain indirectly, together with the mass of the $W$ boson, the mass of the Higgs boson. 
We must notice that the definition of the mass depends on the renormalization scheme used.
With the exception of top mass measurements derived from the production cross section,
experimentally the invariant mass of decay products or the observed kinematics properties are used and may be compared to matrix element based predictions.
Therefore experiments measure the mass as encoded by the MC generators
which is usually related to the mass used for the EWK fits by: $m_{\rm top}^{\rm MC}\approx m_{\rm top}^{\rm \bar{MS}}+10$~GeV/c$^2$.
The dominant systematic uncertainty is the jet energy scale (JES).
Both ATLAS and CMS experiments have carried out an extensive calibration of the JES from di-jet and $Z/\gamma$+jets events,
and have pinned down the overall uncertainty to $\approx$2-3\% depending on the kinematics ($\pt$ and $\eta$) of the jet.
Currently, even in the high pileup regime at which the LHC runs,
the JES uncertainty is mostly dominated by uncertainties from the absolute scale, ISR/FSR, fragmentation
and single particle response in the calorimeter.
In-situ JES measurement can be performed, in particular using the $W\rightarrow qq'$ decays in the lepton+jets channel.
This is a powerful method which allows the experiments to reduce the main uncertainty and focus on a better understanding
of the modelling of signal namely non-perturbative, ISR or jet-parton matching scales and Q$^2$ effects.
It's only limited by the fact that the quarks from $W$ decays are mostly light.

The ATLAS collaboration has performed a combined fit of $m_{\rm top}$
by using templates with varied JES and $m_{\rm top}$~\cite{ATLAS:2012aj}.
Data from the lepton+jets channel are fit with likelihood and 
$m_{top}=$174.4 $\pm$0.6$_{\rm stat}$ $\pm$2.3$_{\rm syst}$ GeV/c$^2$ is measured.
The systematic uncertainty is dominated by flavour-specific JES, ISR/FSR.
Given also the fact that the top quark has color there are 
non-perturbative effects which alter the kinematics properties observed and therefore the mass.
These effects may depend on kinematics such as the production channel, the boost of the top quark or the final state used, among others.
Non-perturbative effects usually known as color reconnection (CR) effects are estimated
from simulation using \textsc{PYTHIA~6.4}~\cite{Skands:2010ak} and are found 
to contribute at the level of $\sim$3\% for the systematic uncertainty associated to this measurement.
The CMS experiment has also performed a combined fit of $m_{\rm top}$ in the lepton+jets channel employing an ideogram technique~\cite{CMS:11015top}. 
After a kinematic fit an event likelihood is defined by taking into account the expected contribution from correct assignments, wrong permutations
and also $\ttbar$ events where at least one of the decay products (jet or lepton) has not been properly reconstructed or selected.
Background events are also taken into account.
The individual event likelihoods are combined in the sample allowing one to extract $m_{\rm top}$ (and JES) from the maximum likelihood found.
The top mass is measured to be:
$m_{\rm top}=$172.6 $\pm$0.6$_{\rm stat}$ $\pm$1.2$_{\rm syst}$ GeV/c$^2$
with the systematic uncertainty being dominated by the modelling of the signal
namely the choice of the Q$^2$ and jet-parton matching scales and by the uncertainty in the flavour-specific JES corrections.
The effects of CR and underlying event tune on this measurement are currently being evaluated by the experiment.

The top quark mass has also been measured in the dilepton channel by the CMS experiment
employing the KIN$_b$ method which solves numerically the equations for the kinematics of the $\ttbar$ pair decay
using different hypothesis for the $\ttbar$ imbalance along the beam line according to the expectations from simulation.
In this case the measurement is dominated by the JES uncertainty followed by the modelling of the signal and
$m_{\rm top}=$173.3 $\pm$1.2$_{\rm stat}$ $\pm$2.6$_{\rm syst}$ GeV/c$^2$ is obtained~\cite{CMS:11016top}.

\begin{figure}
\centering
\includegraphics[height=2.8in]{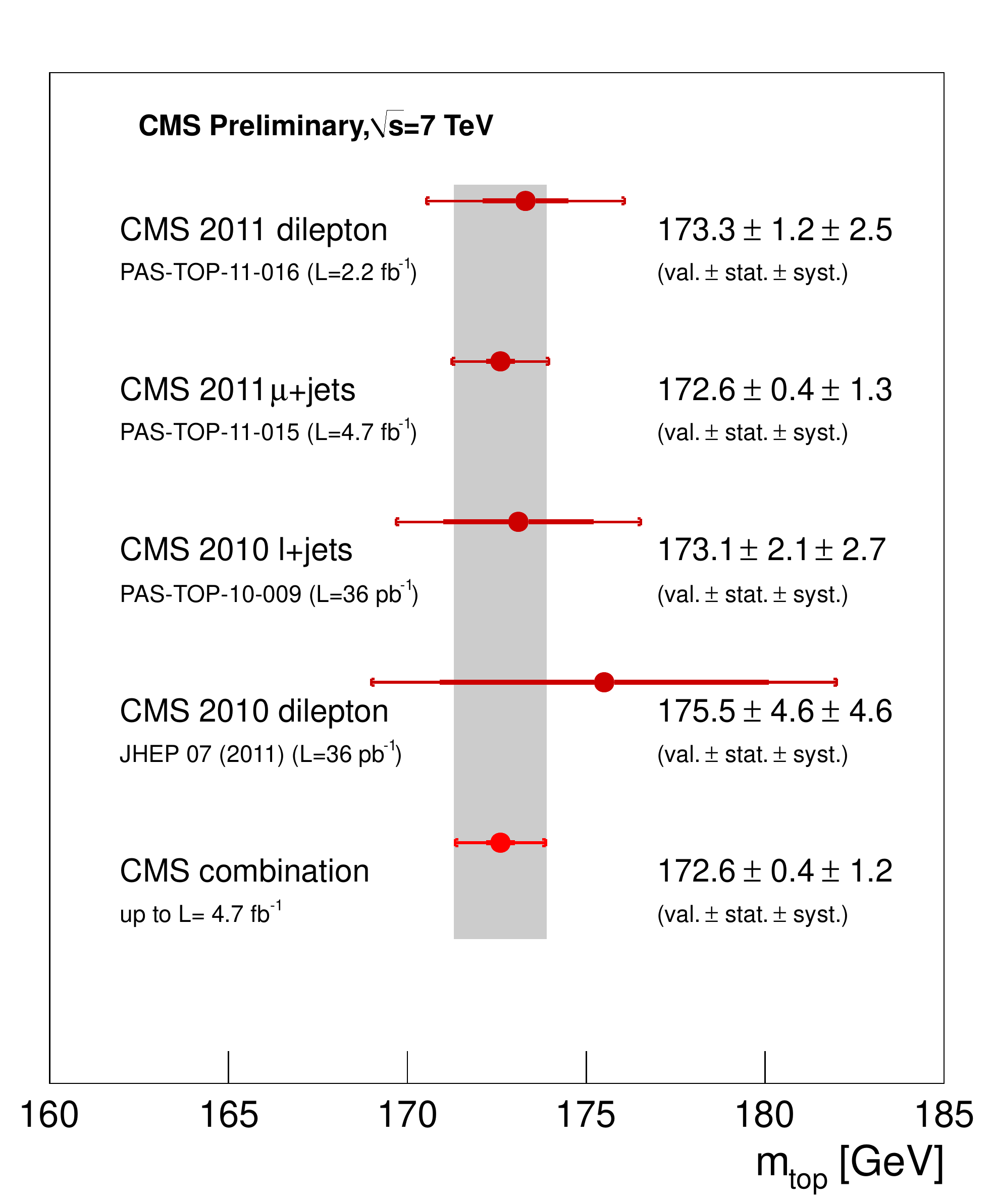}~~~
\includegraphics[height=2.7in]{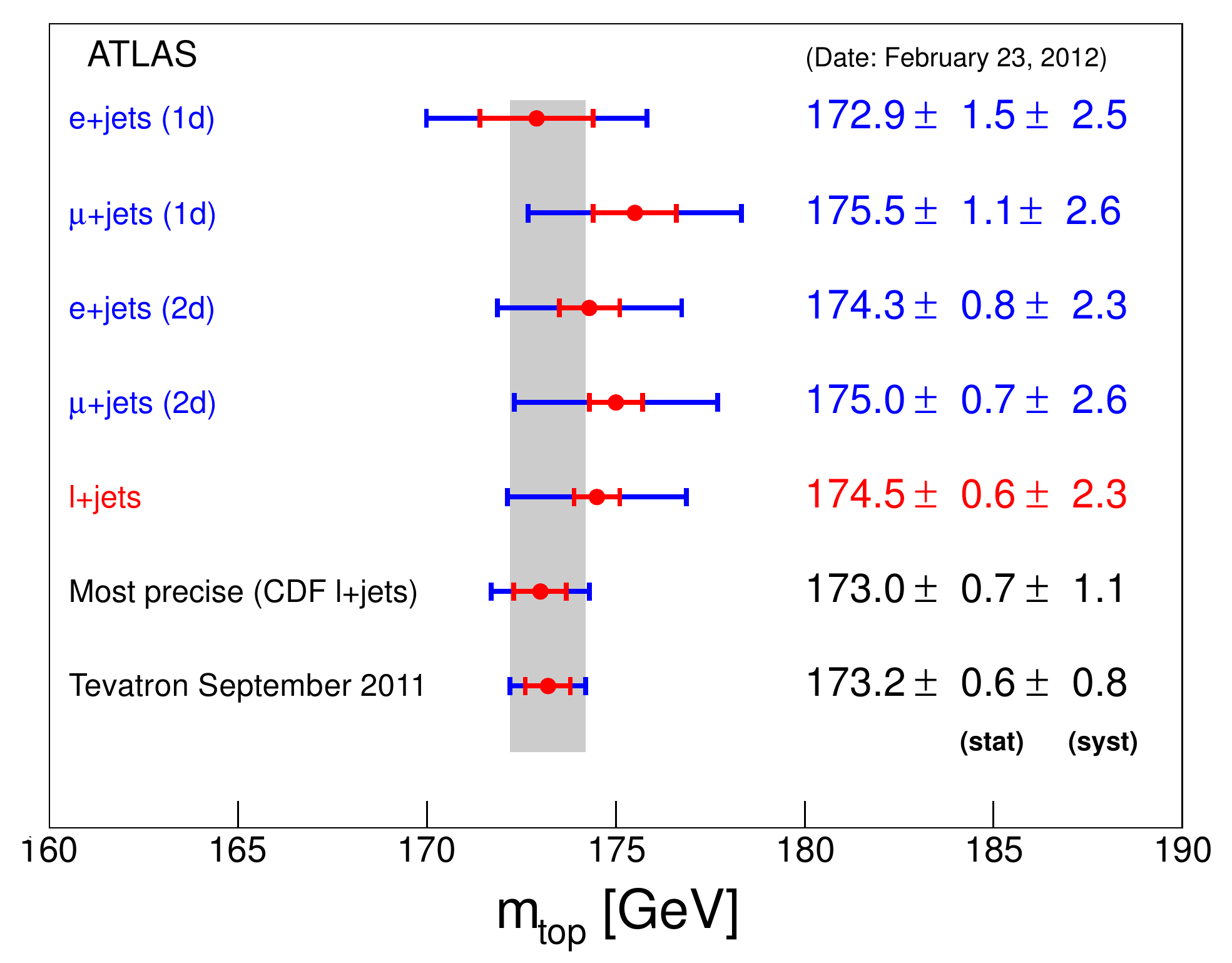}
\caption{Summary of the top mass measurements carried by the ATLAS ({\em left}) and CMS ({\em right}) experiments.}
\label{fig:topmass}
\end{figure}

Figure~\ref{fig:topmass} summarises the different top mass measurements carried out by the LHC experiments.
A good agreement with the Tevatron measurements is found and the overall uncertainty attained by each experiment
starts to be competitive with the most precise measurements~\cite{Lancaster:2011wr}.
A combination of these measurements, once all the uncertainties are estimated 
is expected to yield a significant improvement in the knowledge of $m_{\rm top}$.

The difference between the top/anti-top masses has also been measured at the LHC.
The observation of a deviation from the null result is a signature of CPT violation in the top quark sector.
In this measurement many systematic uncertainties cancel out as a difference between masses is measured.
The CMS collaboration has reconstructed $m_{\rm top}$ from the $\mu^{\pm}$+jets samples 
using a kinematic fit and the combination of jets with lowest $\chi^2$.
The final measurement of the mass in each sample is taken  after combining the event-per event likelihood with the ideogram method.
The result obtained is in good agreement with the SM prediction and it is currently statistically limited:
$\Delta m_{\rm top}=$-0.44 $\pm$0.46$_{\rm stat}$ $\pm$0.27$_{\rm syst}$ GeV/c$^2$~\cite{Chatrchyan:2012ub}.

\section{Summary}
\label{sec:summary}

The LHC experiments have entered the precision era for top quark physics.
The results presented indicate an excellent agreement with the
theoretical predictions in all aspects explored so far.
The challenge laying ahead is the reduction of the systematic
uncertainties either from more precise theoretical computation 
either by constraining some of these uncertainties from data
(e.g. ISR/FSR contribution, $Q^2$ scale, CR effects).
More precise characterisation of the top quark will help to shed light
on the EWKSB mechanism and the role the top quark plays in it.
Precision measurements of the top quark production environment
will be of great value in the continuation of searches for new physics
to which the top constitutes an important background.

%
%
\section*{Acknowledgments}
The author acknowledges the Top quark group conveners from the ATLAS and CMS experiments:
M.J.~Costa, M.~Cristinziani, R.~Chierici and R.~Tenchini for all the help provided in preparing this talk
and the organisers of the ``EW Interactions and Unified Theories'' of the 2012 {\em Rencontres de Moriond}
for the pleasant and warm environment. 

%
%


\section*{References}
\begin{thebibliography}{99}

\bibitem{Aad:2008zzm} 
  The ATLAS Collaboration,``The ATLAS Experiment at the CERN Large Hadron Collider'',
  JINST {\bf 3}, S08003 (2008).
  
\bibitem{Chatrchyan:2008aa} 
  The CMS Collaboration,
  ``The CMS experiment at the CERN LHC'',
  JINST {\bf 3}, S08004 (2008).


\bibitem{bigi} I.~I.~Y.~Bigi,
``On The Hadronization Of Top Quarks'',
\Journal {\PLB} {175} {233} {1986}

\bibitem{atlaspublic} The ATLAS Collaboration, ``ATLAS Experiment: Public Top Results'',
https://twiki.cern.ch/twiki/bin/view/AtlasPublic/TopPublicResults

\bibitem{cmspublic} The CMS Collaboration, ``CMS Top Physics Results'',\\
https://twiki.cern.ch/twiki/bin/view/CMSPublic/PhysicsResultsTOP

\bibitem{ATLAS:11121top}
The ATLAS Collaboration,
``Measurement of the ttbar production cross-section in pp collisions
at $\sqrt{s} =$~7~TeV using kinematic information of lepton+jets events'',
ATLAS-CONF-2011-121

\bibitem{ATLAS:11153top}
The ATLAS Collaboration,
``Measurement of the inclusive t tbar gamma cross section with the ATLAS detector'',
ATLAS-CONF-2011-153''

\bibitem{Sadia} Khalil,~S. ``Top pair cross secion measurement in the  l+jets channel with the CMS detector'', these proceedings

\bibitem{CMS:11013top}
  The CMS Collaboration, 
``Measurement of Top Quark Pair Differential Cross Sections at $\sqrt{s}=$~7 TeV'',
  CMS-PAS-TOP-11-013

\bibitem{ATLAS:11142top}
The ATLAS Collaboration,
``Reconstructed jet multiplicities from the top-quark pair decays and associated jets in pp collisions at $\sqrt{s}=$~7~TeV measured with the ATLAS detector at the LHC'',
ATLAS-CONF-2011-142

\bibitem{ACER}
Kersevan,~B.P.  and Richter-Was¸~E. 
``The Monte Carlo event generator AcerMC version 2.0 with interfaces
to PYTHIA 6.2 and HERWIG 6.5'', 
arXiv:0405247[hep-ph]

\bibitem{PYTHIA}
Sjostrand,~T. and  Mrenna,~S. and Skands,~P.,
``PYTHIA 6.4 physics and manual'', 
arXiv:0603175[hep-ph]

\bibitem{CMS:11021top}
The CMS Collaboration,
``Measurement of the single top t-channel cross section in pp collisions at $\sqrt{s}=$~7~TeV''
CMS-PAS-TOP-11-021

\bibitem{CMS:11022top}
The CMS Collaboration,
``Search for single top production in the tW-channel'',
CMS-PAS-TOP-11-022

\bibitem{Collaboration:2012ux} 
  The ATLAS Collaboration,
  ``Measurement of the t-channel single top-quark production cross section in pp collisions at sqrt(s) = 7 TeV with the ATLAS detector'',
  arXiv:1205.3130 [hep-ex]

\bibitem{ATLAS:11118top}
The ATLAS Collaboration,
``Search for s-Channel Single Top-Quark Production in pp Collisions at $\sqrt{s} =$~7~TeV'',
ATLAS-CONF-2011-118

\bibitem{ATLAS:11104top}
The ATLAS Collaboration,
``Search for Wt associated production in dilepton final states with 0.70 fb$^{-1}$ of 7 TeV pp collision data in ATLAS'',
ATLAS-CONF-2011-104

\bibitem{PhysRevD.83.091503}
Kidonakis,~N.,
``Next-to-next-to-leading-order collinear and soft gluon corrections for $t$-channel single top quark production'',
\Journal{\PRD}{83}{091503}{2011}

\bibitem{PhysRevD.81.054028}
Kidonakis,~N.,
``Next-to-next-to-leading logarithm resummation for $s$-channel single top quark production'',
\Journal{\PRD}{81}{054028}{2010}

\bibitem{Kidonakis:2010ux} 
  Kidonakis,~N.
  ``Two-loop soft anomalous dimensions for single top quark associated production with a W- or H-,''
\Journal{\PRD}{82}{054018}{2010}

\bibitem{Nedden} Nedden,~M. ``Single top results with the ATLAS detector'', these proceedings

\bibitem{ATLAS:2012ao} 
  The ATLAS Collaboration,
  ``Observation of spin correlation in $t \bar{t}$ events from pp collisions at sqrt(s) = 7 TeV using the ATLAS detector'',
  arXiv:1203.4081 [hep-ex].

\bibitem{ATLAS:11141top}
The ATLAS Collaboration,
``Measurement of the top quark charge in pp collisions at $\sqrt{s}=$7~TeV in the ATLAS experiment'',
ATLAS-CONF-2011-141

\bibitem{CMS:11031top}
  The CMS Collaboration, ``Constraints on the Top-Quark Charge from Top-Pair Events'',
  CMS-PAS-TOP-11-031

\bibitem{CMS:11028top}
  The CMS Collaboration, ``Search for t to Zq'',
  CMS-PAS-TOP-11-028

\bibitem{ATLAS:11061top}
  The ATLAS Collaboration,
  ``Search for FCNC top quark processes at 7 TeV with the ATLAS detector'',
  ATLAS-CONF-2011-061

\bibitem{CMS:11029top}
  The CMS Collaboration,
 ``First measurement of $B(t\rightarrow Wb)/B(t\rightarrow Wq)$ in the dilepton channel in pp collisions at $\sqrt(s)=$7~TeV'',
  CMS-PAS-TOP-11-029

\bibitem{Czarnecki:2010wh}
  Czarnecki,~A. and. Korner,~J.G. and Piclum,~J. H.,
 ``Helicity fractions of W bosons from top quark decays at NNLO in QCD'', 
 \Journal {\PRD} {81} {111503} {2010}

\bibitem{CMS:11020top}
  The CMS Collaboration, ``W helicity in top pair events'',
  CMS-PAS-TOP-11-020

\bibitem{protos}
Aguilar-Saavedra~J. A., et al., 
``Probing anomalous Wtb couplings in top pair decays,''
 Eur. Phys. J. C 50 (2007) 519

\bibitem{Collaboration:2012ky} 
  The ATLAS Collaboration,
  ``Measurement of the W boson polarization in top quark decays with the ATLAS detector'',
  arXiv:1205.2484 [hep-ex]


\bibitem{PhysRevD.84.054017}
Gabrielli,~E. and Raidal,~M.,
``Effective axial-vector coupling of gluon as an explanation to the top quark asymmetry'',
\Journal {\PRD} {8} {054017} {2011}

\bibitem{CMS:11030top} 
   The CMS Collaboration,
  ``Differential measurements of the charge asymmetry in top quark pair production'',
   CMS-PAS-TOP-11-030

\bibitem{Rodrigo}
Rodrigo,~G. ``The ttbar asymmetry in the SM'', these proceedings


\bibitem{ATLAS:2012an} 
  The ATLAS Collaboration,
  ``Measurement of the charge asymmetry in top quark pair production in pp collisions at sqrt(s) = 7 TeV using the ATLAS detector'',
  arXiv:1203.4211 [hep-ex].

\bibitem{Skands:2010ak} Skands,~P.Z., ``Tuning Monte Carlo Generators: The Perugia Tunes'', \Journal{\PRD}{82}{074018}{2010}, arXiv:1005.3457 [hep-ph]

\bibitem{ATLAS:2012aj} 
  The ATLAS Collaboration,
  ``Measurement of the top quark mass with the template method in the $\ttbar\rightarrow$  lepton + jets channel using ATLAS data'',
  arXiv:1203.5755 [hep-ex]

\bibitem{CMS:11015top}
  The CMS Collaboration, ``Measurement of the top quark mass in the muon+jets channel'',
  CMS-PAS-TOP-11-015

\bibitem{CMS:11016top}
  The CMS Collaboration, ``Measurement of the top quark mass in the dilepton channel in pp collisions at $\sqrt{s}$=7TeV'',
  CMS-PAS-TOP-11-016

\bibitem{Lancaster:2011wr} 
  [Tevatron Electroweak Working Group and for the CDF and D0 Collaborations],
  ``Combination of CDF and D0 results on the mass of the top quark using up to 5.8~fb$^{-1}$ of data'',
  arXiv:1107.5255 [hep-ex].

\bibitem{Chatrchyan:2012ub} 
  The CMS Collaboration,
  ``Measurement of the mass difference between top and antitop quarks'',
  arXiv:1204.2807 [hep-ex]


\end{thebibliography}
\end{document}